\documentclass[]{aastex631}

\graphicspath{{./}{figures/}}
\usepackage{amsmath}
\usepackage{soul}
\usepackage{subfigure}
\usepackage{multirow}

\begin{document}

\title{Solar Wind Heating Near the Sun: A Radial Evolution Approach}

\author[0000-0001-6018-9018]{Yogesh}
\affiliation{Department of Physics and Astronomy, University of Iowa, Iowa City IA 54224, USA}
\affiliation{NASA Goddard Space Flight Center, Greenbelt, MD, 20771, USA}
\affiliation{The Catholic University of America, Washington, DC 20064, USA}

\author[0000-0003-0602-6693]{Leon Ofman}
\affiliation{NASA Goddard Space Flight Center, Greenbelt, MD, 20771, USA}
\affiliation{The Catholic University of America, Washington, DC 20064, USA}
\affiliation{Visiting, Tel Aviv University, Tel Aviv, Israel}

\author[0000-0001-6038-1923]{Kristopher Klein}
\affiliation{Lunar and Planetary Laboratory, University of Arizona, Tucson, AZ 85721, USA}

\author[0000-0002-8941-3463]{Niranjana Shankarappa}
\affiliation{Lunar and Planetary Laboratory, University of Arizona, Tucson, AZ 85721, USA}

\author[0000-0002-7365-0472]{Mihailo Martinovi\'{c}}
\affiliation{Lunar and Planetary Laboratory, University of Arizona, Tucson, AZ 85721, USA}

\author[0000-0003-1749-2665]{Gregory G. Howes}
\affiliation{Department of Physics and Astronomy, University of Iowa, Iowa City IA 54224, USA}

\author[0000-0002-3808-3580]{Parisa Mostafavi}
\affiliation{Johns Hopkins Applied Physics Laboratory, 
Laurel, MD 20723, USA}

\author[0000-0002-5240-044X]{Scott A Boardsen}
\affiliation{Goddard Planetary Heliophysics Institute, University of Maryland, Baltimore County, MD 21250, USA}
\affiliation{NASA Goddard Space Flight Center, 
Greenbelt, MD 20771, USA}

\author[0000-0002-4001-1295]{Viacheslav M Sadykov}
\affiliation{Physics \& Astronomy Department, 
Georgia State University, 
Atlanta, GA 30303, USA}

\author[0000-0002-6302-438X]{Sanchita Pal}
\affiliation{Centre for Space Science and Technology, Indian Institute of Technology Roorkee, Uttarakhand, 247667, India} 
\affiliation{NASA Goddard Space Flight Center, Greenbelt, MD, 20771, USA}

\author[0000-0002-6849-5527]{Lan K Jian}
\affiliation{NASA Goddard Space Flight Center, 
Greenbelt, MD 20771, USA}

\author[0009-0000-9917-2694]{Aakash Gupta}
\affiliation{Physical Research Laboratory, Navrangpura, Ahmedabad 380009, India} 
\affiliation{Indian Institute of Technology-Gandhinagar, Gandhinagar 382055, India}

\author[0000-0003-2693-5325]{D. Chakrabarty}
\affiliation{Physical Research Laboratory, Navrangpura, Ahmedabad 380009, India}

\author[0000-0001-6673-3432]{B.\ L.\ Alterman}
\affiliation{NASA Goddard Space Flight Center, Greenbelt, MD, 20771, USA}

\author[0000-0003-1138-652X]{Jaye L Verniero}
\affiliation{Heliophysics Science Division,
NASA Goddard Space Flight Center,
Greenbelt, MD 20771, USA}

\author[0000-0002-5699-090X]{K. W. Paulson}
\affiliation{Smithsonian Astrophysical Observatory, Cambridge, MA 02138, USA}

\author[0000-0002-9954-4707]{Jia Huang}
\affiliation{Space Sciences Laboratory, University of California, Berkeley, CA 94720-7450, USA}

\author[0000-0002-0396-0547]{Roberto Livi}
\affiliation{Space Sciences Laboratory, University of California, Berkeley, CA 94720-7450, USA}

\author[0000-0001-5030-6030]{Davin E. Larson}
\affiliation{Space Sciences Laboratory, University of California, Berkeley, CA 94720-7450, USA}

\author[0000-0001-6868-4152]{Christian Möstl}
\affiliation{Austrian Space Weather Office, GeoSphere Austria, Graz, Austria}

\author[0000-0001-9992-8471]{Emma E. Davies}
\affiliation{Austrian Space Weather Office, GeoSphere Austria, Graz, Austria}

\author[0009-0004-8761-3789]{Eva Weiler}
\affiliation{Austrian Space Weather Office, GeoSphere Austria, Graz, Austria}
\affiliation{Institute of Physics, University of Graz, Graz, Austria}

\begin{abstract}

Characterizing the plasma state in the near-Sun environment is essential to constrain the mechanisms that heat and accelerate the solar wind.
In this study, we use Parker Solar Probe (PSP) observations from Encounters 1 through 24 to investigate the radial evolution of solar wind plasma and magnetic field properties in this region. 
Using intervals with high field-of-view ($>$85\%) coverage, we derive the radial profiles of magnetic field strength ($|B|$), proton density ($N$), bulk speed ($V$), total proton temperature ($T$), parallel ($T_\parallel$) and perpendicular ($T_\perp$) temperatures, temperature anisotropy ($T_\perp/T_\parallel$), plasma beta ($\beta$), Alfv\'{e}n Mach number ($M_A$), and magnetic field fluctuations ($\delta B/B$) for sub and super-Alfv\'{e}nic regions. 
In super-Alfv\'{e}nic regions, power-law of $|B|$, $N$, $V$, and $T$ as a function of heliocentric distance are broadly consistent with previous \textit{Helios} results at $>0.3$ AU. 
The radial evolution of the components of the temperature tensor reveals distinct behavior: $T_\perp$ decreases monotonically with distance, whereas $T_\parallel$ exhibits a non-monotonic trend—decreasing in the sub-Alfv\'{e}nic region, increasing just beyond the Alfv\'{e}n surface. 
We interpret the increase in $T_\parallel$ as a proxy for proton beam occurrence.   
We further examine the evolution of magnetic field fluctuations, finding decreasing radial/parallel fluctuations but enhanced tangential/normal/perpendicular fluctuations in sunward direction. 
These fluctuations may provide free energy for beam generation and particle heating via wave-particle interactions. 
\end{abstract}

\section{Introduction} \label{sec:intro}

The solar wind is a continuous outflow of hot, nearly fully ionized plasma from the solar corona that permeates the heliosphere and mediates the connection between the Sun and interplanetary environment \citep{Parker1958}. 
As it expands outward from the solar corona, the solar wind undergoes substantial radial evolution, manifested in changes to its speed, temperature, density, and associated coherent structures and turbulent fluctuations. 
Understanding this evolution is essential for addressing the heating and acceleration of the solar wind, the formation of large- and small-scale structures, and the transfer of energy between fields and particles in the heliosphere. 

The bulk solar wind consists primarily of protons, $\alpha$ particles, and electrons, with a minor contribution from heavy ions \citep{Marsch2006}. Unlike the other constituents, heavy ions are likely tracers and are not dynamically relevant for the evolution of the solar wind \citep{Alterman2025a, Geiss1995,Geiss1995a,von2000}.
Particle detectors onboard spacecraft are capable of measuring in situ the three-dimensional velocity distribution functions (VDFs) of the dominant species \citep[i.e., protons, $\alpha$ particles, and electrons, etc; ][]{Ogilvie1995,Kasper2016,Durovcova2019,Durovcova2021,Abraham2022, Owen2020}. 
The analyses of Helios particle detector observations provided detailed characterization of the solar wind between 0.3 and 1~AU  \citep{Marsch1982,Durovcova2019}, while Ulysses extended these results to larger heliocentric distances (1.3--5.4~AU) and out of ecliptic plane, offering key insights into the solar wind evolution beyond 1~AU \citep{Neugebauer1996,Goldstein2000,Reisenfeld2001}  and the latitudinal variation of solar wind properties \citep[e.g.,][]{McComas1998,Yogesh2024}.

Early in situ observational studies were limited to heliocentric distances $\geq 0.3$ AU as measurements closer to the Sun were not available. 
The launch of of the Parker Solar Probe \citep[PSP,][]{Fox2016} in 2018 has provided in situ measurements deep into the inner heliosphere ($<70$ solar radii ($R_s$) or $<0.33$ AU), opening a new window for studying the radial evolution of the near-Sun solar wind. 
Recent PSP studies have examined electron VDFs between 0.13 and 0.5~AU, revealing how electron properties evolve in the solar wind \citep{Abraham2022}. They show the radial density evolution of the electron core, halo, and strahl population.  
In addition, \citet{Mostafavi2022, Wang2025, Ran2024} reported the statistical properties of the alpha–proton differential flow close to the Sun. 
The differential velocity increases first for Alfv\'en Mach number, $M_A < 1$ up to $M_A \simeq 2$, and then begins to decrease, suggesting that the alpha particles may preferentially accelerate well above the critical Alfv\'{e}n surface. 

As the solar wind accelerates away from the Sun, it transitions from sub-Alfv\'enic to super-Alfv\'enic speeds, exceeding the speed of plasma waves restored by magnetic tension \citep{Alfven1942}. The boundary between these regimes is known as the Alfv\'en surface \citep{Badman2025, DeForest2014}. This transition may play a key role in the turbulent heating of the solar wind and in the formation of magnetic switchbacks \citep[e.g.,][]{Ruffolo2020,Tafti2024}. In the sub-Alfv\'enic regime, disturbances in the solar wind can generate waves that propagate back toward the Sun, establishing a causal feedback on the outflow \citep{Belcher1971, Verdini2009, Chandran2021}. Understanding the properties of both regions is important for elucidating the physical processes governing the solar wind.

\citet{Safrankova2023} presented the first comprehensive statistical analysis of compressive and non-compressive magnetic field fluctuations in the inner heliosphere, demonstrating that turbulence undergoes rapid evolution.
Such rapid changes in turbulence can serve as important sources of energy that particles can absorb. 
This energy absorption may, in turn, contribute to the formation of particle beams \citep[e.g.,][]{Araneda2008,Maneva2013,Bianco2025}. 
Taken together, these results underscore that the region below $\sim 0.3$~AU plays a critical role in shaping both the kinetic and fluid properties of the solar wind. 

Understanding the radial evolution of these properties is a key aspect in identifying which heating and acceleration mechanisms \citep{Rivera2024, Alterman25b,Parashar2026, Kasper2019,Parker1958} are active at different heliocentric distances. 
Recently, the radial distribution of key parameters, including solar wind speed, magnetic field strength, and the number density and temperature of protons and alpha particles were studied by \cite{Liu2024}. 
Their findings revealed that previous radial models based on Helios and other earlier missions often fail to align with PSP observations in the near-Sun region. 
However, \cite{Jia2022AGU, Maruca2023, Brown2025} find consistent results after some inter-calibration between multiple spacecraft from the inner heliosphere out to beyond the orbit of Pluto.

Although only a few studies have examined the radial evolution of solar wind parameters using PSP observations, these efforts are often limited to a small number of PSP encounters or rely on partial moments from the SPAN-I instrument. Furthermore, the radial evolution of the solar wind in the sub-Alfv\'enic and super-Alfv\'enic regions has not yet been studied in detail. In this paper, we investigate the radial evolution of the solar wind in both of these regimes.

The Solar Probe Analyzer for Ions (SPAN-I) instrument has a restricted field of view (FOV) due to obstruction by the PSP thermal heat shield \citep{Livi2022}, which can limit the accuracy of derived plasma moments, especially higher order moments such as the temperature tensor. 
In this study, we explicitly account for the SPAN-I FOV and include only VDFs that are resolved, ensuring that the analyzed plasma measurements are in FOV and reliable. We are removing data with a systematic bias (limited FOV) due to constraints of the instrument and its spacecraft accommodation.
We incorporate data including the closest perihelia achieved by PSP, spanning Encounters 1-24.
By systematically selecting fully observed VDFs and incorporating the innermost PSP measurements, this work advances our understanding of the inner heliosphere and extends it into the deepest portion of the sub-Alfv\'{e}nic region explored to date, thereby providing new insights into the radial evolution of these plasma properties.

The remainder of this paper continues as follows.
Section~\ref{sec:data} details the data sources used.
Section~\ref{sec:obs} describes the FOV calculation for SPAN-I.
Section~\ref{sec:res} presents the results.
Section~\ref{sec:disc} provides the discussion and conclusions.

\section{Data} \label{sec:data}
The Solar Wind Electrons Alphas and Protons (SWEAP) instrument suite onboard PSP measures ion and electron thermal VDFs in the near-Sun solar wind \citep{Kasper2016}. 
The suite consists of Solar Probe Analyzer for Ions \citep[SPAN-I,][]{Livi2022}, as well as two electron analyzers \citep[SPAN-E,][]{Whittlesey2020} and a Sun-pointing Faraday cup, the Solar Probe Cup \citep[SPC,][]{Case2020}. 
Together, these instruments provide nearly complete sky coverage during most encounters, with only minor gaps in ion measurements \citep{Kasper2016}.

In this study, we focus on SPAN-I measurements. 
SPAN-I is a top-hat electrostatic analyzer with mass discrimination capability, designed to measure three-dimensional ion VDFs over an energy range of 2 eV to 30 keV. 
Its FOV spans $247.5^\circ \times 120^\circ$, and it employs a time-of-flight section to distinguish protons, $\alpha$-particles, and heavier ions. 
A portion of the SPAN-I FOV is occulted by PSP’s thermal shield, with the occultation fraction depending on the magnetic field orientation and solar wind flow angle with respect to the detector near perihelia.
This occultation leads to the measurement of ``partial moments'' of density, velocity, and the temperature tensor in which the derived quantities reflect only the portion of the plasma VDF sampled by the instrument and are therefore systematically incomplete.  
To address this limitation, we follow the procedure in Appendix 2.8 of \citet{Marco2024T} to compute the effective FOV, determining whether a VDF is sufficiently captured and the anisotropic temperature is resolved.

Magnetic field measurements are provided by the FIELDS instrument suite onboard PSP \citep{Bale2016}. 
The FIELDS suite includes fluxgate magnetometers (MAG), with MAG data sampled at a cadence of up to 293 Hz. 
We use down-sampled 4 vectors per cycle data for the present analysis.

We have also removed intervals associated with interplanetary coronal mass ejections (ICMEs) from the data. For this purpose, we use the living ICME catalog compiled by \cite{Mostl2020}, available at \url{https://helioforecast.space/icmecat}.

\section{SPAN-I Field of View}\label{sec:obs}

We use solar wind ion properties derived from SPAN-I measurements.
We use the `\texttt{PSP\_SWP\_SPI\_SF00\_L3\_MOM}' data product. The cadence varies from a few minutes to a few seconds at different distances and different encounters. We have used the data with original cadence. We analyze data spanning from 2018 October 2 to 2025 July 31. The current version of ICMECAT extends only up to 30 April 2025; therefore, ICMEs occurring between 1 May 2025 and 31 July 2025 were not removed from the data set. Although this results in an imperfect temporal overlap, it is not expected to significantly influence our results, as the analysis is based on a long time interval and a large statistical sample. 

These plasma parameters are reliable only when most of the proton distribution is within the instrument FOV. 
To ensure the proper interpretation of the data, a proxy for estimating the proton FOV coverage is required. 
We follow the approach to calculate FOV discussed in \citet{Marco2024T}.  
The approach utilizes two reduced distribution functions in azimuth ($\phi$) and elevation ($\theta$). 
The angular coordinates are defined in the instrument frame. The angular bins are determined by the electrostatic potentials applied to the anodes and deflectors of the electrostatic analyzer. This binning scheme is illustrated in Figure~2 of \citet{Livi2022}.
One is summed over the front anode bins (in the sunward direction), yielding counts as a function of azimuth. The other is summed over all deflection angles, yielding count distributions as a function of elevation.
These summed counts, converted into differential energy flux, are included in the SPAN-I level-3 data product. 

\begin{figure}
\begin{center}
\includegraphics[width=0.95\textwidth]{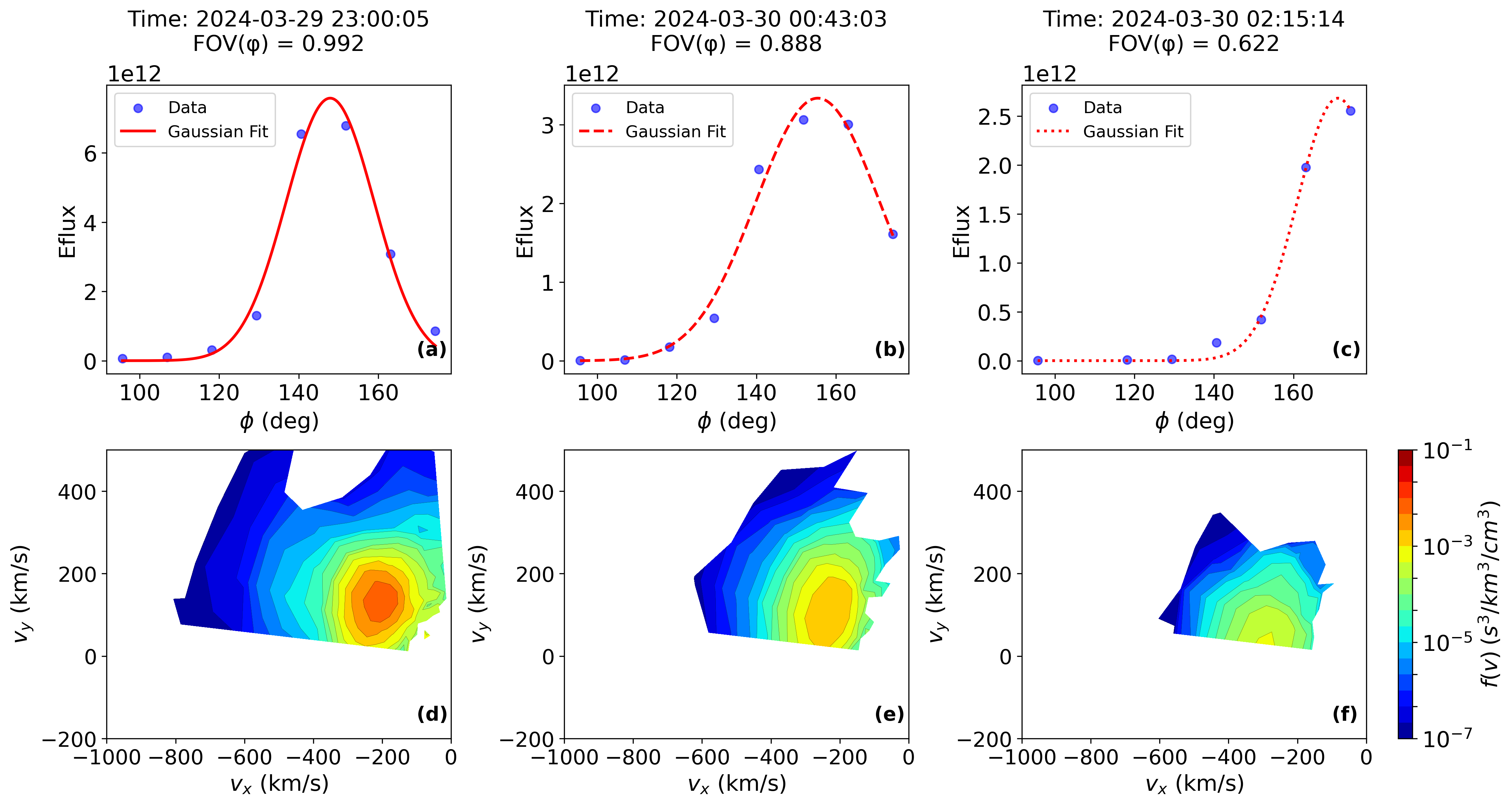}

\caption{Example of FOV calculation for the time indicated by the white vertical solid, dashed and dotted lines in Figure \ref{fig:fov_energy}. 
Panels (a) to (c) show an example of 0.99 (99\%), 0.89 (89\%) and 0.62 (62\%) FOV coverage. 
The lower three panels (d)-(f) show the associated VDF slices in the $v_x$-$v_y$ ($\phi$) plane. 
The 1D Gaussian fits are  also displayed for three cases. 
The effect of the occultation is evident in the VDFs that are cut off in their lower part.}
\label{fig:fov_ex}
\end{center}
\end{figure}

To calculate the FOV, we fit the summed differential energy flux ($J_E$) as a function of each angle to a one-dimensional Gaussian function. 
This fitting requires that at least four data points to be non-zero. 
The Gaussian is expressed as  
\begin{equation}
f(x) = h \exp \left[ - \frac{(x - x_0)^2}{w^2} \right],
\label{eq:gaussian_fit}
\end{equation}
where $h$ is the peak amplitude, $x_0$ is the center, and $w$ is the characteristic width ($\sqrt{2}\,\sigma$) of the distribution. 
Figure~\ref{fig:fov_ex} illustrates three examples of Gaussian fit for azimuthal angles ($\phi$) within the instrument FOV. 

To evaluate the FOV coverage, we compute the integrated area of the fitted Gaussian within
the angular limits of the instrument, defined as $\theta_1 < \theta < \theta_2$ and
$\phi_1 < \phi < \phi_2$. The values of $\theta_1$ and $\theta_2$ are $-51.776^\circ$ and $52.471^\circ$, respectively, while $\phi_1$ and $\phi_2$ are $174.375^\circ$ and $95.625^\circ$.
 
The integration is carried out using the cumulative distribution function (CDF) of the Gaussian,
 \begin{equation}
\mathrm{CDF}(z) = \frac{1}{2} \left[ 1 + \mathrm{erf}\left( \frac{z}{\sqrt{2}} \right) \right],
\label{eq:cdf}
\end{equation}
where $\mathrm{erf}$ is the error function, and the normalized coordinates $z$ are defined as  
\begin{align}
z_{\phi_1} &= \frac{\phi_1 - \phi_0}{w_\phi}, &
z_{\phi_2} &= \frac{\phi_2 - \phi_0}{w_\phi}, \\
z_{\theta_1} &= \frac{\theta_1 - \theta_0}{w_\theta}, &
z_{\theta_2} &= \frac{\theta_2 - \theta_0}{w_\theta}.
\end{align}
In these equations, $w_\theta$ and $w_\phi$ are the widths of the Gaussian functions for the $\theta$ and $\phi$ fits, respectively.
The fractional coverage in each direction is then obtained from the difference of the CDF evaluated at the boundaries:
\begin{align}
A_\phi &= \mathrm{CDF}(z_{\phi_2}) - \mathrm{CDF}(z_{\phi_1}), \\
A_\theta &= \mathrm{CDF}(z_{\theta_2}) - \mathrm{CDF}(z_{\theta_1}).
\end{align}
Finally, the combined FOV coverage is given by the product of the two fractional coverages:
\begin{equation}
A_{\mathrm{FOV}} = A_\phi \, A_\theta.
\end{equation}

This method of calculating the SPAN-I FOV is primarily effective for determining whether the proton core is fully captured within the instrument’s view. 
However, it does not address whether proton beams or secondary proton populations are within the FOV. 
This approach in principle works for alphas, but it requires more sophisticated approach that is beyond the scope of this paper.
The inclusion of these proton beam and alpha properties will be explored in future studies using two-dimensional summed VDFs, enabling a more comprehensive characterization of the solar wind velocity distribution.

\begin{figure}[htbp]
\begin{center}
\includegraphics[width=0.95\textwidth]{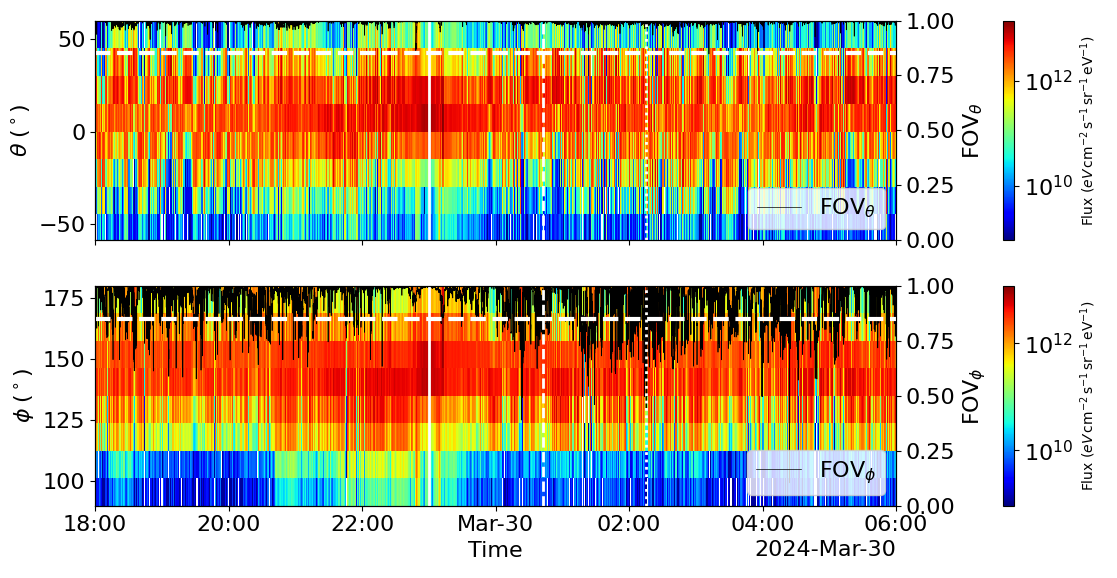}
\caption{Energy Flux and FOV observations from PSP Encounter 19. 
The upper and lower panels display the variation of energy flux with respect to the $\theta$ and $\phi$ angles, respectively. 
The color bar represents the energy flux, while the right y-axis (in black) indicates the FOV. The white dashed horizontal line marks the 0.85 (85\%) FOV in both panels. 
The vertical solid, dashed and dotted white line corresponds to the FOV calculation illustrated in Figure \ref{fig:fov_ex}.}
\label{fig:fov_energy}
\end{center}
\end{figure}

Figure~\ref{fig:fov_ex} illustrates the differential energy flux as a function of $\phi$  observed by SPAN-I on 2024/03/29 at 23:00:05~UT, 2024/03/30 at 00:43:03~UT and 2024/03/30 at 02:15:14~UT. 
Panels (a) through (c) of Figure~\ref{fig:fov_ex} show examples of 0.99, 0.88 and 0.62 FOV coverage, and the lower three panels (d)-(f) show the associated VDF slices in the $v_x$-$v_y$ ($\phi$) plane. 
Comparing the panels from left to right, we observe that the resulting VDF becomes progressively less well defined as a larger portion of the VDF’s azimuthal angle is outside of the SPAN-i FOV.
In the most extreme case, with only 62\% coverage, the peak of the full VDF is not fully resolved, even though the 1D fit identifies a clear maximum. 
This suggests that while the 1D fit enforces the presence of a peak, the complete 3D measurement does not confirm that the VDF is fully captured. 
Therefore, it is essential to use data with a high enough FOV coverage to ensure accurate characterization of the VDF.

The VDFs can also be recovered using the Gyrotropic Slepian Reconstruction \citep{Das2025, Das2026}. The Gyrotropic Slepian Reconstruction method fits Slepian functions and uses them to reconstruct the partial VDFs observed by PSP. This approach provides a means to recover VDFs from incomplete measurements. In the present study, however, we restrict our analysis to intervals with good FOV coverage. Our approach is based purely on direct observations. 

Figure~\ref{fig:fov_energy} illustrates an example from the perihelion of Encounter~19, which occurred on 2024/03/30 at 02:21~UT.
The upper panel shows the variation of differential energy flux as a function of the $\theta$ and time. 
The right $y$-axis (black curve) indicates the corresponding FOV coverage in $\theta$. 
The horizontal dashed white line marks the threshold at $FOV_\theta = 0.85$. 
The threshold is chosen based on analysis shown in Appendix~\ref{ap:fov0.85}. 
Similarly, the lower panel presents the variation of differential energy flux with respect to the azimuthal angle ($\phi$). 
The black curve denotes the fractional FOV coverage in $\phi$, while the dashed white horizontal line highlights the reference level at $FOV_\phi = 0.85$. 
The vertical white lines correspond to the three cases at 2024/03/29 23:00:05~UT, 2024/03/30 00:43:03~UT and 2024/03/30 02:15:14~UT shown in Figure~\ref{fig:fov_ex}.
When the black line drops below the reference white FOV level, we consider the VDF to be outside of the SPAN-I FOV.
In Figure~\ref{fig:fov_energy}, this is more common for the $\phi$ direction than $\theta$ because of the obstruction caused by heat shield.

\begin{figure}[htbp]
\begin{center}
\includegraphics[width=0.8\textwidth]{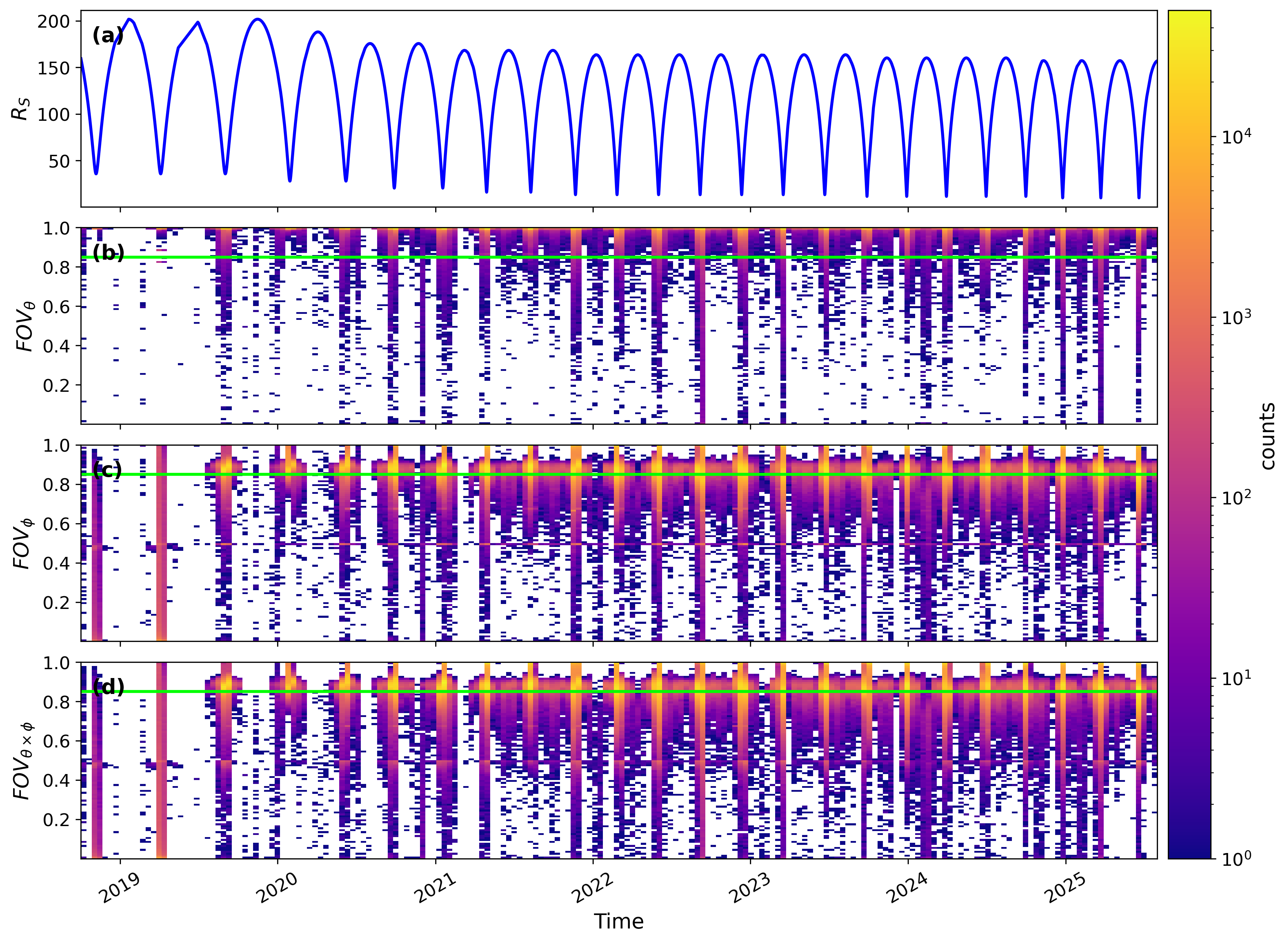}
\caption{The variation of SPAN-I FOV for Encounters 1–24. 
Panels (a)–(d) display the radial distance from the Sun ($R_s$), the FOV for $\theta$, the FOV for $\phi$, and the combined FOV ($\theta \times \phi$), respectively. 
The green line in the lower three panels shows the FOV=0.85. 
The color bar represent the number of observations. }
\label{fig:fov_var}
\end{center}
\end{figure}

This procedure is applied to proton measurements from SPAN-I for PSP Encounters 1 through 24 to quantify the degree of FOV coverage. 
Figure~\ref{fig:fov_var} shows the variation of the FOV during these encounters as a function of radial distance, separately for $\theta$, $\phi$, and the combined case ($\theta \times \phi$). 
It can be seen that $FOV_{\theta}$ remains close to unity for a large fraction of the observations, whereas $FOV_{\phi}$ approaches unity primarily near perihelia. 
The combined FOV shows similar variation to $FOV_{\phi}$.

\begin{figure}[htbp]
\begin{center}
\includegraphics[width=0.95\textwidth]{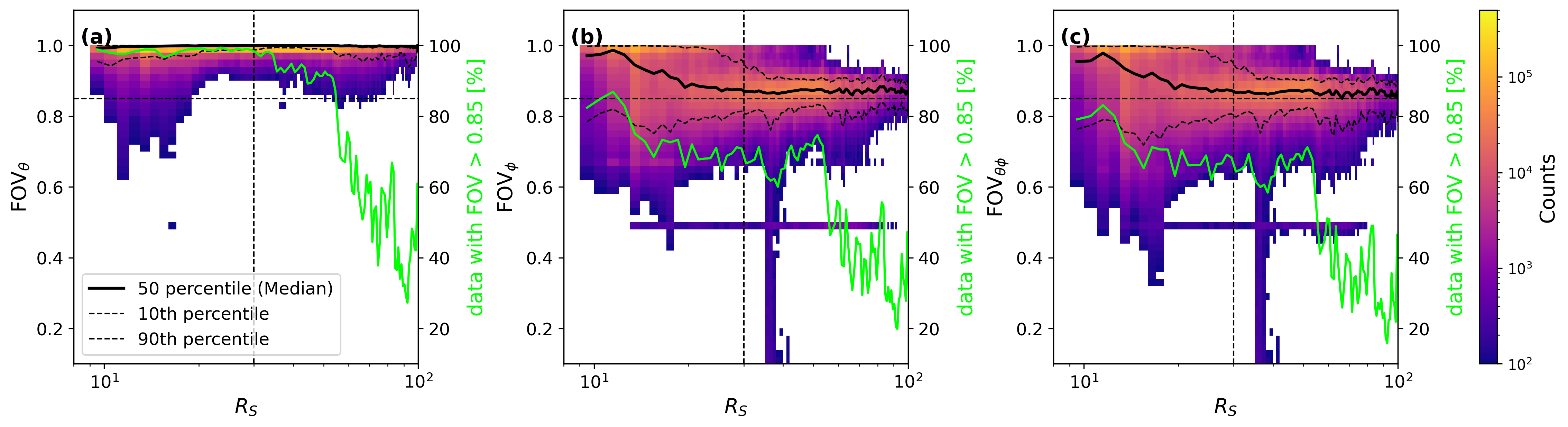}
\caption{The variation of SPAN-I FOV during Encounters 1 - 24 with radial distance. 
The FOV for $\theta$, the FOV for $\phi$, and the combined FOV ($\theta \times \phi$) are shown. 
The black dashed horizontal line shows the FOV=0.85. 
The median values are shown with black curve. 
The 10\% and  90\% levels are shown by black dashed lines. 
The lime line based on the right y-axis shows that  percentage of data having FOV$>$0.85.
The dashed vertical line shows $R_s$=30.  }
\label{fig:fov_var_rad}
\end{center}
\end{figure}

Figure~\ref{fig:fov_var_rad} shows the radial variation of the FOV in the $\theta$ and $\phi$ directions, as well as for the combined case ($\theta \times \phi$). 
The solid black line represents the median FOV for the analyzed intervals.
The lime green curve indicates the percentage of total measurements (including those not fitted) with $FOV > 0.85$.
It should not be confused that the black line represents the median of all data; rather, it represents the median of observations for which the FOV could be calculated.
For $FOV_\theta$, the coverage remains high — typically exceeding $85\%$ — within $R \lesssim 30R_s$. Beyond approximately $35R_s$, the quality of the FOV systematically decreases, as indicated by the declining fraction of data with $FOV_\theta > 0.85$. 
The $\theta$ direction is not obstructed by the heat shield but is partially blocked by the spacecraft structure, leading to a gradual reduction in the effective FOV.
To ensure reliable measurements and avoid regions with reduced coverage in both directions, we adopt a radial cut-off at $R = 30R_s$ and restrict our primary analysis to this inner region. 
This selection ensures that the dataset includes intervals with at least one direction having full FOV coverage.

The $FOV_\phi$ approaches unity closer to the Sun; however, a fraction of VDFs near the Sun still exhibit $FOV_\phi < 0.85$. 
It is important to emphasize that the distributions shown are restricted to observations that satisfy the fitting criterion of at least four data points in both $\theta$ and $\phi$. 
As shown by the lime green curve in Figure~\ref{fig:fov_var_rad}, the fraction of observations with a good FOV relative to the total number of observations decreases with increasing heliocentric distance.
When considering the complete dataset, approximately 56\% of observations have $FOV > 0.85$. 
Thus, near-Sun VDFs with limited FOV coverage are not uncommon and can occur even near perihelion.

Nevertheless, the dataset provides statistically significant coverage to investigate the radial evolution close to the Sun. 
The main limitation arises for the heliocentric distances between 9.9 and 11 $R_s$, where the statistics are based only on data from three Encounters (22-24). There are no data below $\sim$ 9.9 Rs.
This limitation will be reduced with observations from future encounters.

\section{Results}\label{sec:res}
\subsection{Radial evolution of solar wind parameters}

\begin{figure}[htbp]
\begin{center}
\includegraphics[width=0.95\textwidth]{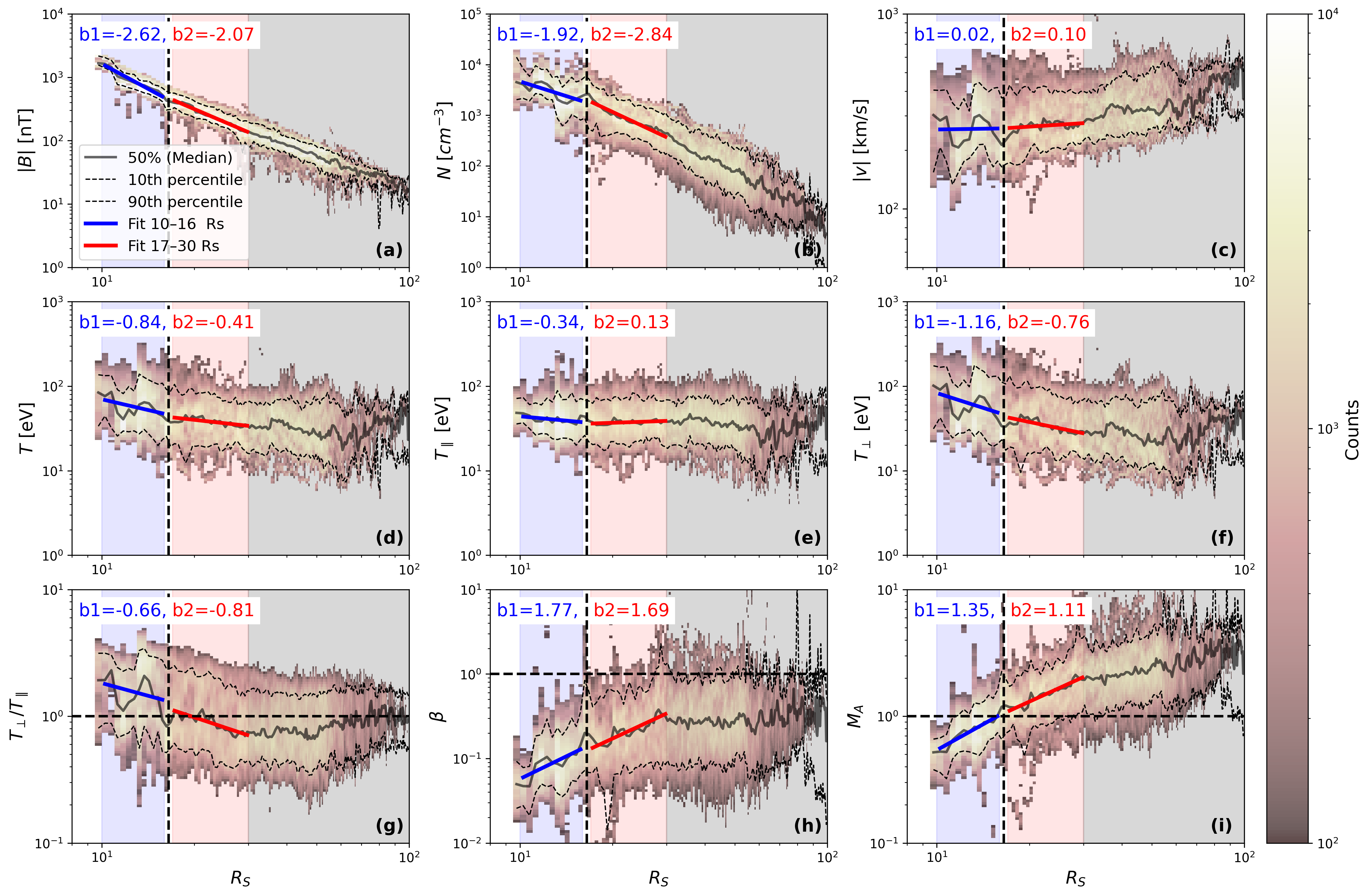}
\caption{Radial evolution of (a) magnetic field strength ($|B|$, nT), (b) proton density ($N$, cm$^{-3}$), (c) proton bulk speed ($|V|$, km s$^{-1}$), (d) total temperature ($T$, eV), (e) parallel temperature ($T_\parallel$, eV), (f) perpendicular temperature ($T_\perp$, eV), (g) temperature anisotropy ($T_\perp/T_\parallel$), (h) plasma beta ($\beta$), and (i) Alfv\'{e}n Mach number ($M_A$). 
The black solid line and the black dashed lines represent the logarithmic median and 10-90 percentile levels, respectively. 
The black dashed vertical line marks the mean Alfv\'{e}n critical surface at $16R_s$ where $M_A = 1$. 
The blue (10–16 $R_s$) and pink shaded region (17–30 $R_s$) indicates the radial range used for the power-law fitting. 
The blue and red lines shows the power-law fits for sub- and super-Alfv\'{e}nic region respectively. }
\label{fig:sw_var}
\end{center}
\end{figure}

Figure~\ref{fig:sw_var} shows the radial profiles of (a) magnetic field strength ($|B|$), (b) proton number density ($N$), (c) proton bulk speed ($|V|$), (d) total proton temperature ($T$), (e) parallel proton temperature ($T_\parallel$), (f) perpendicular proton temperature ($T_\perp$), (g) proton temperature anisotropy ($T_\perp/T_\parallel$), (h) proton plasma beta ($\beta_p$), and (i) Alfv\'{e}n Mach number ($M_A$). 
We use a bin width of 0.5~$R_{\mathrm{s}}$ using the heliocentric distance provided in the Level-3 data set.
In each panel, the black solid line denotes the logarithmic median, and the gray region indicates the 10-90 percentile region.  

The vertical black line at $16.5\,R_s$ corresponds to the median Alfv\'{e}n critical surface, where $M_A \approx 1$, separating the sub-Alfv\'{e}nic and super-Alfv\'{e}nic regimes (see panel i). 
It should be noted that the location of the Alfv\'{e}n surface is not a single value, but exhibits significant variability depending on the solar wind parameters. 
For our analysis, we adopt the statistical median of the Alfv\'{e}n critical surface at $\sim16.5 R_s$, 
which is consistent with the value reported by \citet{Badman2025}.

Interestingly, the plasma beta ($\beta$) profile (panel h) shows a plateau beyond $\sim 30\,R_s$, whereas a monotonic increase with radial distance would be expected. This can also be verified using electron density calculated from quasi thermal noise (QTN) spectra measured by the FIELDS/RFS instrument on PSP \citep{Moncuquet_etal_2020_QTN_PSP, Pulupa_etal_2017_RFS}. 
The comparison can be seen in Figure \ref{fig:sw_qtn} in Appendix~\ref{ap:com_qtn} where plasma beta calculated using electron density and temperature is increasing continuously with heliocentric distance. 
This plateau begins at approximately the same radial distance where the fraction of data with $FOV_\theta > 0.85$ starts to decrease.  
This behavior may arise from instrumental FOV limitations because the plasma instrument was designed to observe the solar wind close to the Sun. 
To mitigate the effect, this region beyond 30 $R_s$ is not included in our radial fitting that follows.

We fit a power-law model separately to the sub-Alfv\'{e}nic ($10$--$16\,R_s$, blue) and super-Alfv\'{e}nic ($17$--$30\,R_s$, pink) regions of the solar wind, while the outer region beyond $30\,R_s$ (gray) is excluded from the fit. 
The lower limit of the sub-Alfv\'{e}nic region is set at $10\,R_s$ to avoid the narrow width of 9.9--10 $R_s$ bin. 
The data in the 9.9--11.4~$R_{\mathrm{s}}$ range are available only from Encounters 22--24. This limited coverage contributes to the larger uncertainties in the power-law fits in the sub-Alfv\'{e}nic region, as shown in Table \ref{tab:sw_var}. These statistics are expected to improve as data from additional encounters become available in the future. To minimize contamination between regimes, we exclude the 16–17~$R_s$ region, where significant overlap between the sub- and super-Alfv\'{e}nic solar wind is expected. We also tested an alternative separation at 16.5~$R_s$ and another approach in which independent power-law fits were performed for the sub- and super-Alfv\'{e}nic regions at similar distances. The resulting fitted exponents remain consistent within the 1--$\sigma$ uncertainties reported in Table~\ref{tab:sw_var}, indicating that our results are robust to the choice of the boundary.

The power-law model is described by the relation
\begin{equation}
Y(R) = a \left(\frac{R}{R_s}\right)^{b},
\end{equation}
where $a$ is the scaling factor at $R = R_s$ and $b$ is the power-law index characterizing the radial variation of the quantity $Y$.

The blue and red lines in Figure~\ref{fig:sw_var} show the resulting best-fit power-law trends. 
The corresponding best-fit parameters and their $1-\sigma$ uncertainties, derived from the covariance matrix, are summarized in Table~\ref{tab:sw_var}. Due to the limited data in the sub-Alfv\'{e}nic region, the uncertainty in the sub-Alfv\'{e}nic region is larger than in the super-Alfv\'{e}nic region.

We have also fit a broken power-law model (not shown here).
The radial profiles of most of the parameters for the broken power-law exhibit a clear break point nearby $16 R_s$, except a few ($|B|$, $T_\perp$ and $T_\perp/T_\parallel$). 
These results indicate a natural separation in the power-law behavior between the sub-Alfv\'enic and super-Alfv\'enic solar wind regions.

The power-law fits reveal distinct radial trends in the plasma parameters across the sub-Alfv\'{e}nic (10--16~$R_s$) and super-Alfv\'{e}nic (17--30~$R_s$) regions. 
In the sub-Alfv\'{e}nic regime, the magnetic field strength decreases rapidly with distance ($b_1 = -2.62 \pm 0.22$), indicating a strong decline in field magnitude close to the Sun. 
The proton density exhibits a dependence aproximatly $r^{-2}$, suggesting a steady state spherical expansion of the plasma. 
The bulk flow speed shows at only a very small slope ($b = 0.02 \pm 0.32$). The 1-$\sigma$ value of power law index (slope) is high because we do not observe any trend in bulk flow speed. The data is also lesser in sub-Alfv\'{e}nic region. The bulk velocity also shows a deep and a hump in the sub-Alfv\'{e}nic region. 

Beyond the Alfv\'{e}n point, in the super-Alfv\'{e}nic range, the magnetic field continues to decrease with an exponent $b_2 = -2.07 \pm 0.08$, indicating nearly spherical expansion. 
This decrease is slightly steeper than previously reported values of around $\sim -1.7$ \citep{Maruca2023}. 
The proton density declines more steeply ($b_2 = -2.84 \pm 0.15$), consistent with ongoing solar wind expansion.
While the proton density decreases with an exponent  below $-2$, the electron density (shown in Figure~\ref{fig:sw_qtn}) exhibits a value close to $-2$, in agreement with \citet{Maruca2023}. The steeper slope in density observed by SPAN-I may be because of the limited FOV as the radial distance decreases.
The bulk velocity increases gradually with $b_2 = 0.10 \pm 0.05$, consistent with similar exponents reported in earlier studies \citep{Maruca2023, Bale2016}. 
 
Overall, the magnetic field strength ($|B|$), proton density ($N$), and temperature ($T$) generally decrease with radial distance, whereas the proton bulk speed ($|V|$) increases slowly in the super-Alfv\'{e}nic region, similar to previously reported results. 
 
The panels (d), (e), (f) and (g) of Figure~\ref{fig:sw_var} illustrate the evolution of $T$, $T_\parallel$, $T_\perp$, and temperature anisotropy ($T_\perp/T_\parallel$). 
In the sub-Alfv\'{e}nic region, the total temperature and its perpendicular component decrease steeply with distance ($b_1 = -0.84 \pm 0.47$  and $b_1 = -1.16 \pm 0.64$, respectively), while the parallel temperature declines more gradually ($b_1 = -0.34 \pm 0.19$). The high 1‑$\sigma$ values of slope in this region are caused by the lower number of solar wind observations at smaller heliocentric distances.
In this study, we use numerical moments and do not segregate out regions of the VDF that may have a beam because that analysis is beyond the scope of our study. As a consequence, enhanced $T_\parallel$ can indicate the presence of unresolved beams.
These temperature variations contribute to a systematic reduction in the temperature anisotropy ($T_\perp/T_\parallel$) with increasing radial distance.
The dip and subsequent hump observed in the total, perpendicular temperatures and anisotropy resemble similar variations seen in the solar wind bulk velocity, which may indicate a common underlying driver that is not investigated further in this study.

In contrast, in the super-Alfv\'{e}nic region, the proton temperature components exhibit significantly smaller exponents. The total and perpendicular temperatures decrease slowly with $b_2 = -0.41 \pm 0.09$ and $b_2 = -0.76 \pm 0.12$, respectively, consistent with previous observations \citep{Jia2020}. 
The parallel temperature shows a slight increase ($b_2 = 0.13 \pm 0.06$), which may be attributed to the enhanced occurrence of field-aligned proton beams.
Observationally, proton beams are less prominent in the sub-Alfv\'enic region and become more abundant in the super-Alfv\'enic solar wind, a feature that has not been widely explored in the literature and that is discussed further in Section~\ref{sec:disc}.
  
Because of these contrasting temperature trends, the anisotropy ($T_\perp/T_\parallel$) does not follow a well-defined radial dependence and cannot be reliably represented by a simple power law. 
Initially, the anisotropy drops toward unity in sub-Alfv\'enic region, then drops below one as $T_\parallel$ increases.  

Panels (h) and (i) show the plasma beta ($\beta$) and the Alfv\'enic Mach number ($M_A$), respectively. Both $\beta$ and $M_A$ increase with radial distance in both regions.
The plasma beta scales as $b_1 = 1.77 \pm 0.39$ in the sub-Alfv\'enic range and remains similar in the super-Alfv\'enic region with $b_2 = 1.69 \pm 0.18$. 
The Alfv\'enic Mach number follows a similar trend, increasing with $b_1 = 1.35 \pm 0.13$ in the sub-Alfv\'enic region and $b_2 = 1.11 \pm 0.08$ in the super-Alfv\'enic region.   


\begin{table}[h!]
\centering
\caption{Power-law fit parameters for sub-Alfv\'{e}nic regime (10--16 $R_s$) and super-Alfv\'{e}nic regime (17--30 $R_s$) distance ranges.}
\begin{tabular}{lcccccc}
\hline
\multirow{2}{*}{Variable} & \multicolumn{2}{c}{sub-Alfv\'{e}nic regime fit } & & \multicolumn{2}{c}{super-Alfv\'{e}nic regime fit } \\
\cline{2-3} \cline{5-6}
 & $a_1 \pm$ $\delta a_1$ & $b_1 \pm$ $\delta b_1$ & & $a_2 \pm$ $\delta a_2$ & $b_2 \pm$ $\delta b_2$ \\
\hline
$|B|$ 
& $(6.71 \pm 3.64)\times10^{5}$ & $-2.62 \pm 0.22$ 
& & $(1.56 \pm 0.38)\times10^{5}$ & $-2.07 \pm 0.08$ \\

$N$ 
& $(3.84 \pm 5.28)\times10^{5}$ & $-1.92 \pm 0.55$ 
& & $(5.81 \pm 2.65)\times10^{6}$ & $-2.84 \pm 0.15$ \\

$|V|$ 
& $240.93 \pm 196.89$ & $0.02 \pm 0.32$ 
& & $193.72 \pm 31.60$ & $0.10 \pm 0.05$ \\

$T$ 
& $489.66 \pm 578.80$ & $-0.84 \pm 0.47$ 
& & $134.58 \pm 35.86$ & $-0.41 \pm 0.09$ \\

$T_{\parallel}$ 
& $96.63 \pm 47.10$ & $-0.34 \pm 0.19$ 
& & $25.25 \pm 4.65$ & $0.13 \pm 0.06$ \\

$T_{\perp}$ 
& $1211.89 \pm 1963.78$ & $-1.16 \pm 0.64$ 
& & $369.50 \pm 133.88$ & $-0.76 \pm 0.12$ \\

$T_{\perp}/T_{\parallel}$ 
& $8.42 \pm 12.11$ & $-0.66 \pm 0.57$ 
& & $11.06 \pm 3.25$ & $-0.81 \pm 0.09$ \\

$\beta$ 
& $(9.72 \pm 10.04)\times10^{-4}$ & $1.77 \pm 0.39$ 
& & $(1.09 \pm 0.63)\times10^{-3}$ & $1.69 \pm 0.18$ \\

$M_A$ 
& $(2.41 \pm 0.79)\times10^{-2}$ & $1.35 \pm 0.13$ 
& & $(4.76 \pm 1.15)\times10^{-2}$ & $1.11 \pm 0.08$ \\

\hline
$\delta |B|/B$ 
& $(6.06 \pm 3.80)\times10^{-4}$ & $0.37 \pm 0.24$ 
& & $(7.50 \pm 2.00)\times10^{-5}$ & $1.03 \pm 0.08$ \\

$\delta B_R/B$ 
& $(2.68 \pm 2.60)\times10^{-3}$ & $0.44 \pm 0.38$ 
& & $(1.57 \pm 0.48)\times10^{-3}$ & $0.58 \pm 0.10$ \\

$\delta B_T/B$ 
& $(4.96 \pm 2.75)\times10^{-2}$ & $-0.18 \pm 0.22$ 
& & $(5.27 \pm 1.41)\times10^{-2}$ & $-0.22 \pm 0.09$ \\

$\delta B_N/B$ 
& $(1.10 \pm 0.61)\times10^{-1}$ & $-0.46 \pm 0.22$ 
& & $(6.18 \pm 1.66)\times10^{-2}$ & $-0.26 \pm 0.09$ \\

$\delta B_{\parallel}/B$ 
& $(2.90 \pm 2.62)\times10^{-3}$ & $0.05 \pm 0.35$ 
& & $(6.81 \pm 2.02)\times10^{-4}$ & $0.51 \pm 0.09$ \\

$\delta B_{\perp}/B$ 
& $(4.83 \pm 2.82)\times10^{-2}$ & $-0.29 \pm 0.23$ 
& & $(3.30 \pm 0.80)\times10^{-2}$ & $-0.17 \pm 0.08$ \\

\hline
\end{tabular}\label{tab:sw_var}
\end{table}

\subsection{Radial evolution of Magnetic field fluctuations}

\begin{figure}[htbp]
\begin{center}
\includegraphics[width=0.95\textwidth]{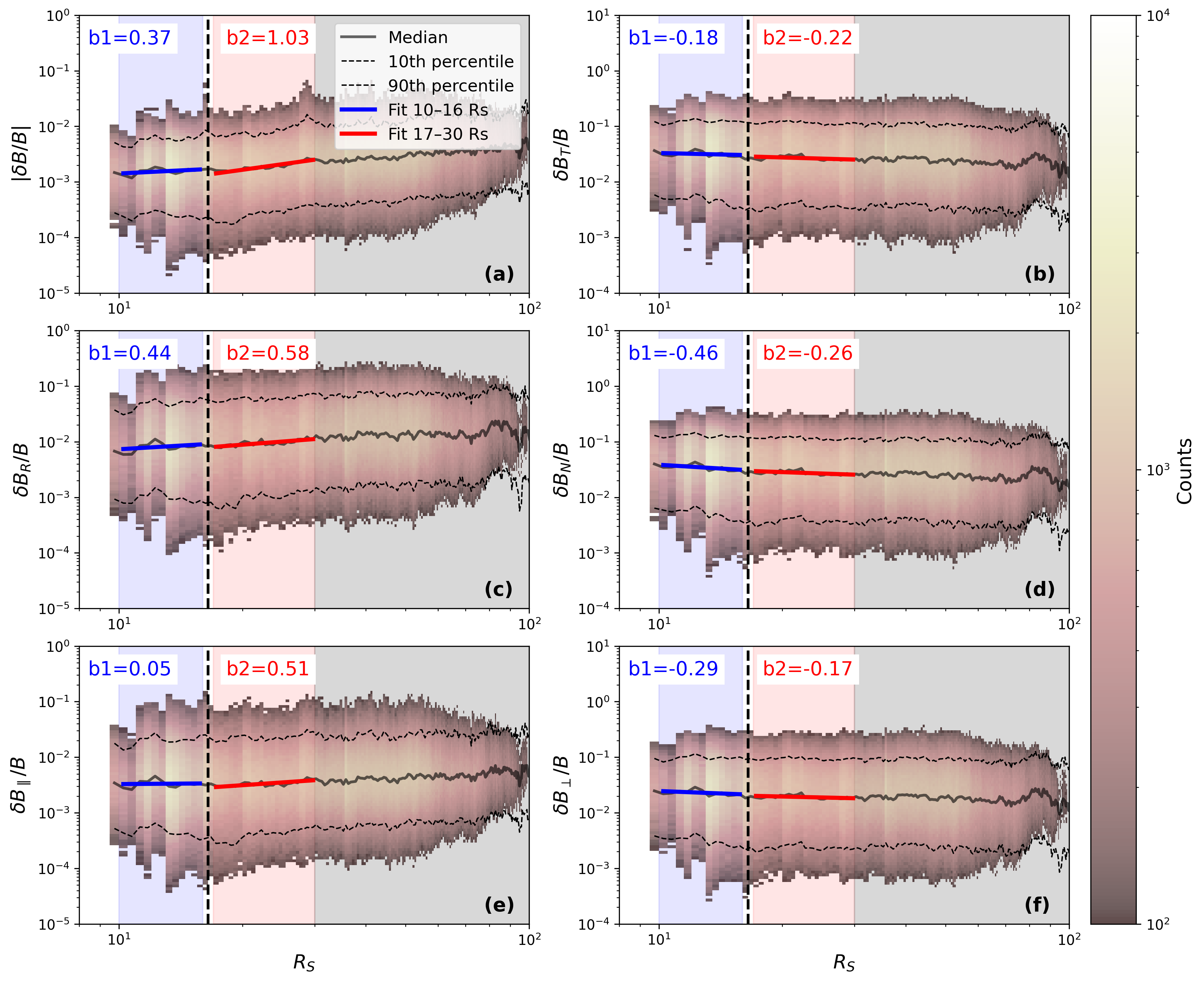}
\caption{Radial evolution of magnetic fluctuations ($|\delta B / B|$) are shown. 
Panels (a), (c) and (e) show the fluctuations in the total, radial and parallel component. Panels (b), (d) and (f) show the fluctuations in tangential, normal and perpendicular components of the magnetic field. 
The black solid line and the black dashed lines represent the median and 10-90 percentile levels, respectively. 
The black dashed vertical line marks the mean Alfv\'{e}n critical surface at $16R_s$ where $M_A = 1$. 
The blue (10–16 $R_s$) and pink shaded region (17–30 $R_s$) indicates the radial range used for the power-law fitting. 
The blue and red lines show the power-law fits for sub- and super- Alfv\'{e}nic regions respectively. 
 }
\label{fig:mag_var}
\end{center}
\end{figure}

To assess the impact of cross-scale transfer and dissipation of energy in the solar wind, we next examine the radial evolution of magnetic field fluctuations. 
Figure~\ref{fig:mag_var} presents the radial evolution of magnetic fluctuations for the total field magnitude, as well as for the radial, tangential, normal, parallel and perpendicular components. Similar trends in the total magnetic field—closely match the behavior of both the radial component and the magnitude of the total field—have also been reported for the first nine orbits of PSP by \cite{Safrankova2023}.
The perpendicular, tangential, and normal fluctuation also shows a trend similar to the perpendicular fluctuation shown in \cite{Safrankova2023}. 

Similar to the plasma parameters, power-law fits are applied to the magnetic field fluctuations over the same sub- and super-Alfv\'{e}nic regions covering  $10$--$16\,R_s$ and $17$--$30\,R_s$, respectively. 
The resulting power-law indices are indicated in each panel of Figure~\ref{fig:mag_var} and summarized in Table~\ref{tab:sw_var}. 
To compute $| \delta B / B |$, a 10s (0.1 Hz) moving average of the magnetic field is subtracted from the instantaneous measurements.
We select a 10-second timescale because it remains within the inertial range even close to the Sun, while encompassing the entire dissipation/kinetic range.
The spectral break frequency varies with heliocentric distance, ranging  approximately 0.5–5 Hz  for the distances under consideration \citep{Safrankova2023}. Tests with different smoothing windows confirm that the overall trends remain consistent, with only the fluctuation amplitudes varying slightly.  
The radial evolution of the normalized magnetic fluctuations, $\delta B/B$, exhibits distinct behaviors in the sub- and super-Alfvénic regimes. In the sub-Alfvénic region, the total, radial, and parallel components show relatively weak power-law dependencies with positive exponent on heliocentric distance, with spectral indices $b_1 = 0.37 \pm 0.24$ (total), $b_1 = 0.44 \pm 0.38$ (radial), and $b_1 = 0.05 \pm 0.35$ (parallel). In contrast, the transverse components exhibit negative slopes, with $b_1 = -0.18 \pm 0.22$ (tangential), $b_1 = -0.46 \pm 0.22$ (normal), and $b_1 = -0.29 \pm 0.23$ (perpendicular). The 1-$\sigma$ values are high in this region due to the limited number of observations at smaller heliocentric distances. These uncertainties are expected to be reduced as more data become available from upcoming encounters; however, in this study we focus on a qualitative comparison. The positive slope in the radial component suggests a gradual enhancement of compressive fluctuations with distance in the sub-Alfvénic region, whereas the negative slopes in the perpendicular components indicate a decay of non-compressive magnetic fluctuations as the solar wind expands.

In the super-Alfvénic region, the total magnetic fluctuation displays a comparatively stronger increase with heliocentric distance, characterized by a spectral index of $b_2 = 1.03 \pm 0.08$, while the radial and parallel components show moderate positive slopes with $b_2 = 0.58 \pm 0.10$ and $b_2 = 0.51 \pm 0.09$, respectively. In contrast, the transverse components continue to exhibit negative power-law indices, with $b_2 = -0.22 \pm 0.09$ (tangential), $b_2 = -0.26 \pm 0.09$ (normal) and $b_2 = -0.17 \pm 0.08$ (perpendicular), indicating a persistent decay of perpendicular magnetic fluctuations with distance.
These results indicate that perpendicular magnetic field fluctuations lose energy more rapidly in the sub-Alfv\'{e}nic region than in the super-Alfv\'{e}nic solar wind. 

From Wentzel--Kramers--Brillouin (WKB) theory, which does not account for dissipation, and from modeling studies \citep{Matteini2024}, the power-law exponents of $\delta B / B$ and $\delta B_\perp / B$ are expected to be 1 and $-0.5$, respectively. For the super-Alfv\'enic solar wind, we find a comparable exponent of 1.03 for $\delta B / B$, while the radial and tangential components exhibit lower exponents than the perpendicular magnetic-field fluctuations. Figure~6 of \cite{Boardsen2015} shows that $\delta B / B$ decreases with radial distance following a power law with an exponent of 0.82, which lies between the exponents of the total and radial magnetic-field components reported in this study. The analysis shown in \cite{Boardsen2015} focused on heliocentric distances from 0.3 to 0.7~AU using \textit{MESSENGER} observations.

\section{Discussion and Conclusions} \label{sec:disc}

The statistical radial evolution of the solar wind provides key observational constraints needed to identify the mechanisms responsible for energy exchange between plasma particles and magnetic fields.
Several physical processes contribute to this energy transfer, 
most notably kinetic mechanisms involving wave--particle interactions 
\citep{Yogesh2025, Ofman2025}. 
These include Landau damping, transit-time damping, stochastic heating, 
and cyclotron-resonant damping \citep{Howes2024}.
Landau damping and transit-time damping energize particles parallel to the magnetic field, whereas cyclotron damping and stochastic heating predominantly energize particles in the perpendicular direction. 
Therefore, examining the radial evolution of different temperature components together with magnetic field fluctuations in the solar wind can help diagnose which heating mechanisms contribute or dominate in different heliospheric regions.

To understand this, we have examined the radial variation of several key solar wind parameters. 
The radial variations of $|B|$, bulk flow proton density ($N$), velocity ($V$), and temperature ($T$) are well studied in the literature farther from the Sun ($\geq0.3$ AU) \citep{Liu2023,Maruca2023}. 
Using Parker Solar Probe observations near perihelion, we present, to the best of our knowledge, the first comparison of the radial evolution of the super- and sub-Alfv\'enic solar wind regions.
In addition, we have analyzed derived parameters such as the temperature components ($T_\parallel$, $T_\perp$), anisotropy ($T_\perp/T_\parallel$), plasma beta ($\beta$), and Alfv\'{e}n Mach number ($M_A$). 
The most important results of this study concern the radial evolution of the proton temperatures and their anisotropy close to the Sun.  

In Figure~\ref{fig:sw_var}, the second row shows the radial evolution of the total temperature $T$, parallel temperature $T_\parallel$, and perpendicular temperature $T_\perp$. 
While $T_\perp$ and the total $T$ exhibit similar decreasing trends, $T_\parallel$ displays a more complex profile: an initial decrease in the sub-Alfv\'{e}nic region, followed by an increase just beyond the Alfv\'{e}n point, and then a subsequent decrease at larger heliocentric distances \citep{Jia2020}.
The power-law fitting applied in the super-Alfv\'{e}nic region of increasing $T_\parallel$ yields a positive exponent, consistent with this rising trend. 
Increasing $T_{\parallel}$ can indicate the emerging presence of a beam population, because proton beams travel along the magnetic field. 
If the beam is not separated from the core in the fit of the distribution or the calculation of the moment, it will artificially increase $T_{\parallel}$. 

The first decreasing trend in $T_\parallel$ occurs in the sub-Alfv\'{e}nic region. 
\cite{Niranjana_2025} shows a drop in the normalized proton heat flux at these distances, which suggests that the proton beams disappear in the sub-Alfv\'enic region.
This region exhibits a steep decrease in the perpendicular magnetic field fluctuation compared to the super-Alfv\'{e}nic wind as shown in figure \ref{fig:mag_var}. 
One possible explanation for the perpendicular heating is stochastic heating \citep{Martinovic2020,Chandran2010}, which has been suggested to be important in the sub-Alfv\'{e}nic regime \citep{Bowen2025}.
This steeper decrease may indicate that the magnetic fluctuations are absorbed providing the energy source of the solar wind plasma heating. 
However, the mechanism by which this process preferentially enhances perpendicular heating requires further investigation.  
The ion temperature anisotropy then drops below unity in super-Alfv\'{e}nic solar wind, most likely due to enhanced $T_\parallel$ from beam generation, and finally approaches unity at larger distances, consistent with solar wind thermalization \citep{Maruca2013, Mostafavi_etal_2024_collision, Johnson_etal_2024, Jagarlamudi2025}. Coulomb collisions can also thermalize the solar wind \citep{Alterman2018}, leading to fewer observations of beams as the plasma moves closer to L1.

Magnetic field fluctuations may provide the key to understanding the generation of beams. 
Due to the small Parker spiral angle below 0.3 AU,  the solar wind flow can be assumed to be purely radial.
Under this assumption, the radial magnetic field component can be considered approximately parallel component, while the tangential and normal components represent perpendicular directions.
Although this mapping is not exact, the observed radial trends of these components remain consistent with such an interpretation which can be seen in figure \ref{fig:mag_var}. 
Our results show that perpendicular fluctuations increase closer to the Sun, whereas parallel fluctuations decrease. It can be seen in Figure \ref{fig:mag_var} that the radial gradients of the magnetic fluctuations in the perpendicular direction are steeper in the sub-Alfv\'{e}nic region compared to the super-Alfv\'{e}nic region.
This behavior suggests a stronger release of fluctuation power in the sub-Alfvénic region, consistent with enhanced wave activity and turbulent energy transfer closer to the Sun. 
The occurrence of ion-scale waves also rises with decreasing heliocentric distance \citep{Liu2023}, possibly linked to these enhanced fluctuations. 
Such fluctuations or strong waves may play a role in beam generation \citep{Araneda2007,Maneva2013,Maneva2015,Maneva2018}. So, this tells us that the beams are probably generated by the waves near sun around the $M_A=1$.
\citet{Niranjana_2025} report a change in the median proton heat flux across the Alfv\'en surface, providing indirect evidence for beam evolution similar to our hypothesis.
Another potential mechanism for beam formation involves drifting alpha particles \citep[e.g.,][]{Ofman2014,Ofman2017}. The alpha–proton differential speed increases closer to the Sun \citep{Mostafavi2022,Wang2025}. 
This enhanced drift can provide a source of free energy, leading to the formation of proton beams that can lead to solar wind anisotropic heating \citep[e.g.,][]{Ofman2017,Ofman2022,Ofman2025}. 
Such alpha particle studies remain an important subject of future study.  

The combined evidence from the power-law fits to plasma parameters and magnetic field fluctuation suggests that the physical processes governing solar wind acceleration and heating in the sub-Alfv\'{e}nic region differ fundamentally from those in the super-Alfv\'{e}nic solar wind, in agreement with \citet{Mostafavi2025}. 
In particular, the stronger radial evolution of both thermal and magnetic properties in the sub-Alfv\'{e}nic regime highlights the importance of ongoing energy exchange close to the Sun. 
These steep radial gradients of thermal and magnetic fluctuations in the sub-Alfv\'{e}nic regimes may indicate significant ongoing energy exchange between particles and fields, which is possibly related to solar wind heating and acceleration, close to the Sun. This interpretation is consistent with previous hypothesis that the near-Sun plasma is subject to distinct acceleration and heating processes compared to the mature, super-Alfv\'{e}nic solar wind \citep[such as,][]{Kasper2019, Holmes2024}. 

In conclusion, the radial evolution of solar wind parameters measured by PSP reveals distinct behaviors in the sub- and super-Alfv\'{e}nic regions. 
Proton temperatures and their anisotropy evolve strongly close to the Sun, likely influenced by perpendicular heating mechanisms such as cyclotron resonance or stochastic heating. 
Magnetic field fluctuations show enhanced perpendicular power near the Sun, providing free energy for heating and beam generation.
These beams cause an increase in parallel temperature and decrease in anisotropy in super-Alfv\'{e}nic regions.
Our findings highlight that solar wind  heating is governed by different physical processes in the sub-Alfv\'{e}nic region compared to the mature, super-Alfv\'{e}nic solar wind. 
Future work involving detailed kinetic modeling and full proton velocity distribution analysis will further clarify the roles of various heating mechanisms and beam generation in the near-Sun solar wind.

\section*{Acknowledgments} 
We thank the PSP mission team for generating the data and making them publicly available. The authors L.O., Y., S.A.B., L.K.J., P.M., V.S., K.G.K., M.M. acknowledge support by NASA grant 80NSSC24K0724. L.O., S.A.B., V.S.  and Y. acknowledge the support by NSF grant AGS-2300961.
Y. and L.K.J. acknowledge support by NASA Heliophysics Guest Investigator Grant 80NSSC23K0447.
G.G.H. acknowledges the support by NASA 80NSSC24K124. Y. acknowledge the support by the College of Liberal Arts and Sciences at the University of Iowa.
K.G.K. acknowledges partial support from NASA contract NNN06AA01C and grant 80NSSC24K0171.
B.L.A. acknowledges Parker Solar Probe funding at NASA/GSFC.
P.M. acknowledges the partial support by the NSF SHINE grant 2401162 and the NASA HGIO grant 80NSSC23K0419. J.H. acknowledges the support by NASA grant 80NSSC23K0737. 
The authors acknowledge CNES (Centre National de Etudes Spatiales), CNRS (Centre National de la Recherche Scientifique), the Observatoire de PARIS, NASA and the FIELDS/RFS team for their support to the PSP/SQTN data production, and the CDPP (Centre de Donnees de la Physique des Plasmas) for their archiving and provision.
The ICMECAT living catalog is available online for the research community at \url{https://helioforecast.space/icmecat} and the filesharing platform figshare, with the latest version available at \url{https://doi.org/10.6084/m9.figshare.6356420}. For this study, the ICMECAT v2.3, updated on 2025 October 15, version 24 on figshare, has been used. This work is supported by ERC grant (HELIO4CAST, 10.3030/101042188). Funded by the European Union. Views and opinions expressed are however those of the author(s) only and do not necessarily reflect those of the European Union or the European Research Council Executive Agency. Neither the European Union nor the granting authority can be held responsible for them.

\appendix

\section{Determination of 85\% FOV limit}\label{ap:fov0.85}

\begin{figure}[htbp]
\begin{center}
\includegraphics[width=0.6\textwidth]{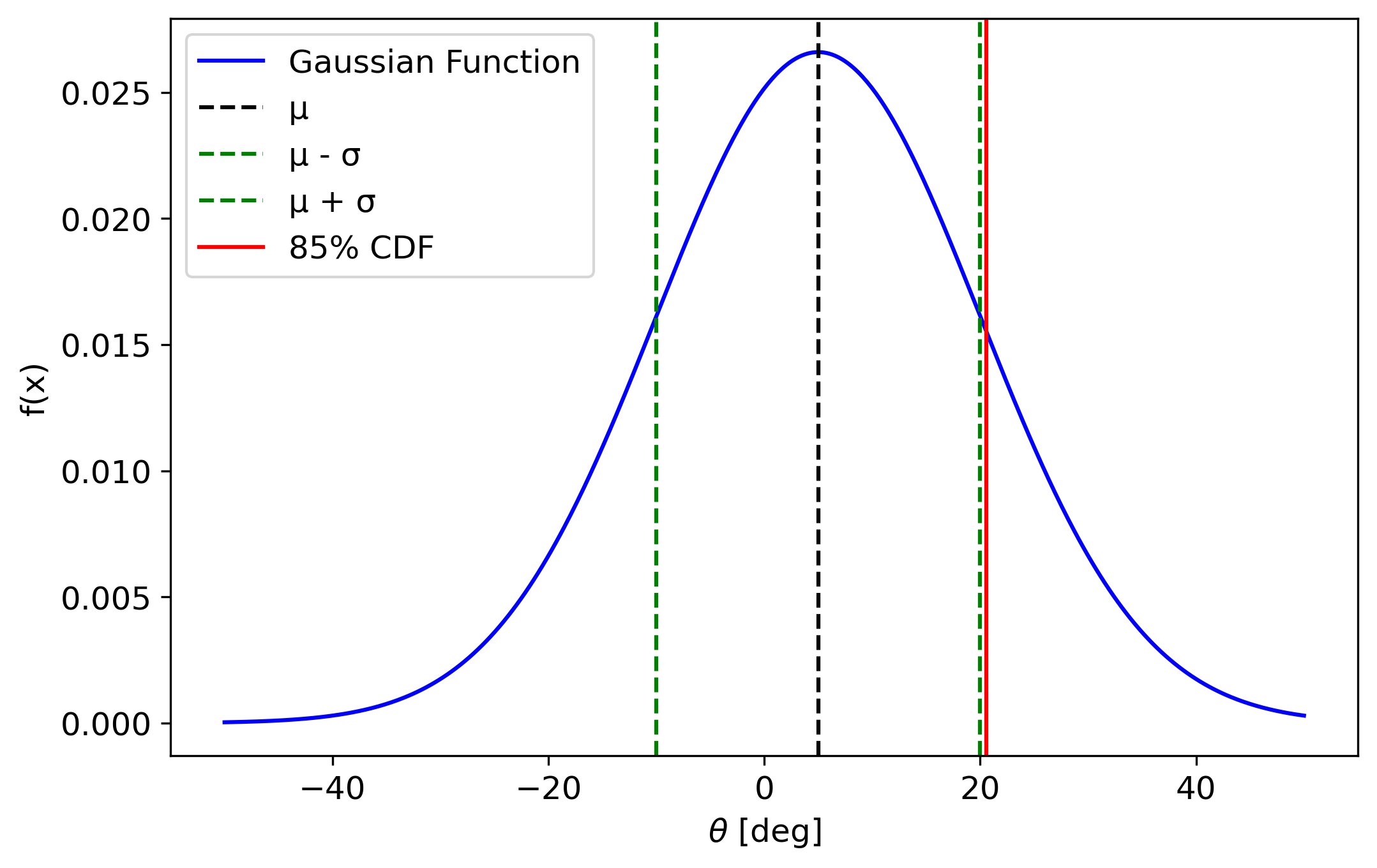}
\caption{The black, green, and red lines represent the center ($\mu$), the 1-sigma ($\sigma$), and the 85\% cumulative distribution, respectively.}
\label{fig:fov0.85}
\end{center}
\end{figure}

A normalized Gaussian distribution as a function of $\theta$ is shown in Figure~\ref{fig:fov0.85}. This distribution is similar to that observed by PSP and is comparable to Figure~\ref{fig:fov_ex}. The mean value, $\mu = 5^\circ$, is indicated by the black dashed line, while the 1-sigma ($\sigma = 15^\circ$) range is marked by the green dashed lines. To quantify field-of-view (FOV) coverage, the cumulative distribution function (CDF) is calculated, the red line representing the fraction of the distribution contained within the FOV. In particular, the 85\% CDF level is highlighted to illustrate effective coverage.  

Since temperature estimates are primarily based on the 1-$\sigma$ width, it is important to examine how this relates to the chosen FOV. As seen in Figure~\ref{fig:fov0.85}, the 85\% CDF encompasses the 1-$\sigma$ region, indicating that selecting 85\% of the FOV is sufficient to capture the bulk of the distribution. This ensures that the derived plasma moments remain representative, although small numerical errors may persist because moments are computed over the entire distribution. Nevertheless, restricting the analysis to 85\% of the FOV provides reliable and robust moment estimates.

\section{Comparing solar wind densities from QTN  and SPAN-I data} 

\begin{figure}[htbp]
\begin{center}
\includegraphics[width=0.95\textwidth]{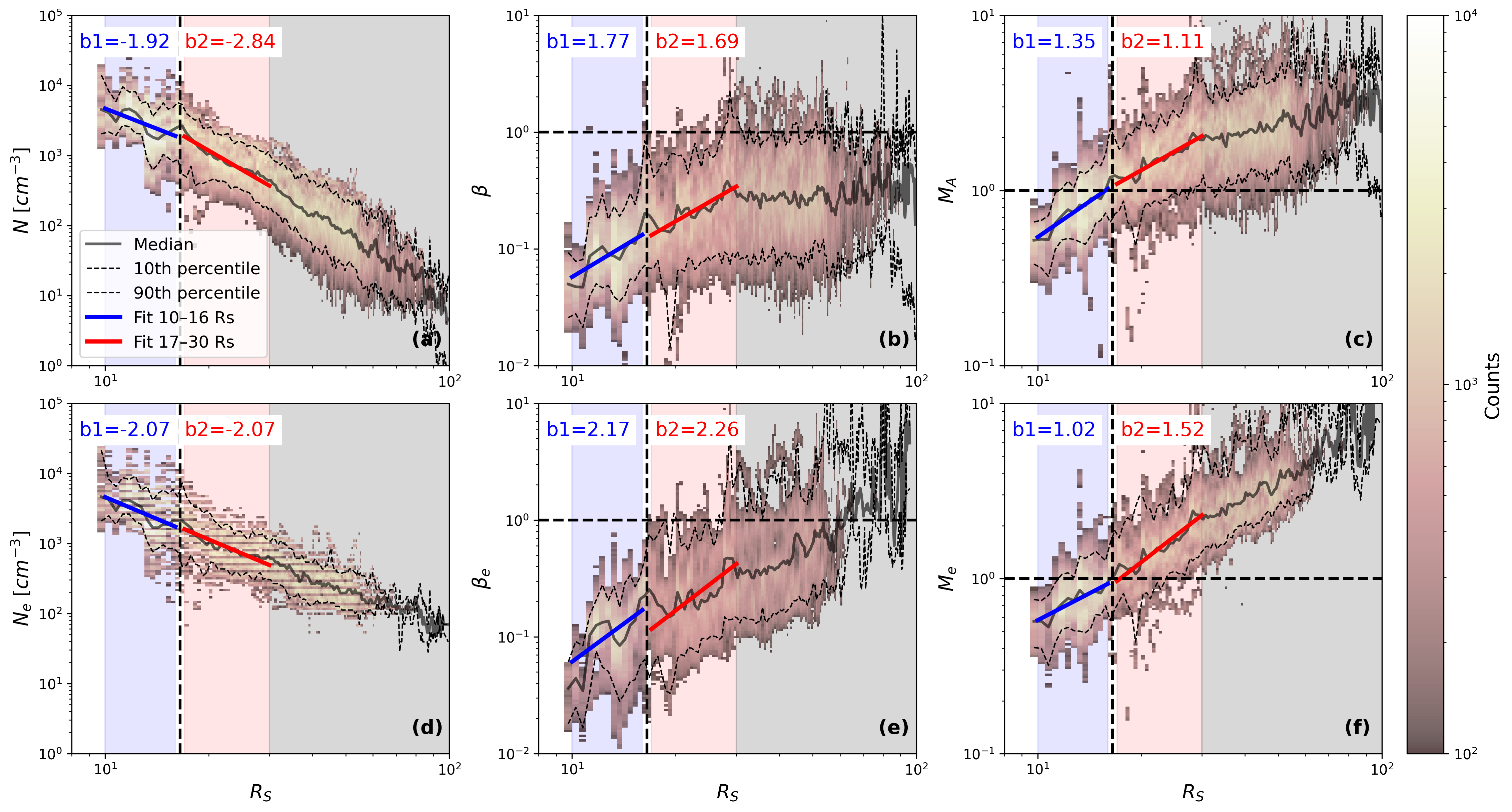}
\caption{Radial evolution of (a) proton density ($N$, cm$^{-3}$), (b) plasma beta ($\beta$), and (c) Alfv\'{e}n Mach number ($M_A$), (d) Electron number density ($N_e$, cm$^{-3}$), (e) plasma beta using electron parameters ($\beta_e$), and (f) Alfv\'{e}n Mach number using electron density ($M_e$). 
The black solid line and the black dashed lines represent the median and 10-90 percentile levels, respectively. 
The black dashed vertical line marks the mean Alfv\'{e}n critical surface at $16R_s$ where $M_A = 1$. 
The blue (10–16 $R_s$) and pink shaded region (17–30 $R_s$) indicates the radial range used for the power-law fitting. 
The blue and red lines shows the power-law fits for sub and super Alfv\'{e}nic region respectively. }
\label{fig:sw_qtn}
\end{center}
\end{figure}

In this section, we compare SPAN-I observations with Quasi-Thermal Noise (QTN) data measured by the FIELDS/RFS instrument on PSP \citep{Moncuquet_etal_2020_QTN_PSP, Pulupa_etal_2017_RFS}.  
We use the proton density from SPAN-I and the electron density from QTN to examine the radial evolution of key plasma parameters.  
Figure~\ref{fig:sw_qtn}, panels (a)–(c), show the proton number density ($N$), plasma beta ($\beta$), and Alfv\'{e}n Mach number ($M_A$) calculated using SPAN-I proton density, whereas panels (d)–(f) present the electron number density ($N_e$), plasma beta ($\beta_e$), and Mach number ($M_e$) derived from QTN electron density and temperature.  

Both the proton and electron densities exhibit a power-law trend with an exponent close to $-2$ in the sub-Alfv\'{e}nic region.  
However, in the super-Alfv\'{e}nic regime, the SPAN-I--derived density decreases much more steeply than the electron density. Quasineutrality of the plasma implies that ion density and electron density should be the same.  
Disagreement between the density measurements can be due to instrumental limitations, such as the limited FOV of SPAN-I.  
This motivates the gray region beyond $R>30~R_s$, where we stop fitting the SPAN-I measurements.  

The plasma beta (see equation~\ref{eq:beta}) from QTN and SPAN-I also shows a difference in the sub-Alfv\'{e}nic power-law exponents (2.17 and 1.77, for QTN and SPAN-I respectively).  
In contrast, $\beta_e$ derived from QTN displays a sharp increase in the super-Alfv\'{e}nic region, whereas $\beta$ calculated from SPAN-I saturates.  

\begin{equation} \label{eq:beta}
\text{Proton plasma beta from SPAN-I: } 
\beta = \frac{2 \mu_0 n_p T_p}{B^2}, \qquad
\text{Electron plasma beta from QTN: } 
\beta_e = \frac{2 \mu_0 n_e T_e}{B^2}.
\end{equation}

\label{ap:com_qtn}

Similarly, the Mach number evolution from the two datasets shows only a modest difference in the sub-Alfv\'{e}nic power-law indices (1.02 for QTN and 1.35 for SPAN-I).  
As with plasma beta, the QTN-derived Mach number exhibits a noticeable jump in the super-Alfv\'{e}nic region.  

A particularly interesting feature is the continuous increase in the plasma beta derived from QTN beyond $\sim 30~R_\odot$, whereas the SPAN-I--based $\beta$ values exhibit a plateau.  
This contrast suggests a limitation of the SPAN-I observations in the super-Alfv\'{e}nic regime, as measurements beyond 30~$R_s$ have limited FOV coverage.  
A similar divergence appears in the Mach number profiles derived from QTN and SPAN-I data.  
Therefore, our imposed upper limit in the radial evolution analysis within the super-Alfv\'{e}nic region is well justified.

The SPAN-E instrument also provides electron density and temperature parameters derived from direct measurements of the electron velocity distribution function. When adequate field-of-view (FOV) coverage is available, SPAN-E serves as a valuable source of electron density estimates. However, during many intervals, the electron moments derived from SPAN-E are affected by limited or incomplete FOV coverage, which can lead to an underestimation of the electron density.
Figure \ref{fig:fov_ex} illustrates a related field-of-view limitation for the ion detector, where ions arriving within certain incoming angles do not reach the detector. For electrons, the situation is further complicated at low energies: positive spacecraft charging repels a fraction of the ambient electrons, while simultaneously enhancing the observed population of photoelectrons. These effects introduce additional challenges in reliably deriving electron moments from SPAN-E data. A detailed discussion of these instrumental and geometric limitations can be found in \cite{McGinnis2021}.
To obtain a robust and geometry-independent estimate of the electron density, we therefore adopt the quasi-thermal noise (QTN)–derived electron density throughout this study. QTN measurements are not subject to particle field-of-view limitations and provide a more reliable density estimate under these observing conditions.

\bibliography{ref}{}
\bibliographystyle{aasjournal}

\end{document}